\documentclass[11pt]{article}

\usepackage[a4paper,margin=1in]{geometry}
\usepackage{graphicx}
\usepackage{amsmath,amssymb}
\usepackage{natbib}
\usepackage{color}
\usepackage{authblk}

\title{Combined thermographic measurement and heat-flux compensation methods for aerodynamic heating evaluation in hypersonic flight}

\author[1]{Kento Inokuma}
\author[1]{Aiko Yakeno}
\author[1]{Yoshiyuki Watanabe}
\author[1]{Kiyonobu Ohtani}

\affil[1]{Institute of Fluid Science, Tohoku University,\\
	2-1-1 Katahira, Aoba-ku, Sendai 980-8577, Miyagi, Japan\\
	Email: inokuma@tohoku.ac.jp}

\date{} 

\begin{document}
	
	\maketitle
	
\begin{abstract}
Novel thermographic measurement and heat-flux compensation methods combined for evaluating aerodynamic heating in hypersonic flight were developed using high-speed thermography.
A hypersonic spherical projectile with a diameter of 8 mm was launched at approximately Mach 5 in the test section of a ballistic range.
Shadowgraph imaging was conducted to visualize the flight trajectory and the shock layer. 
Thermographic measurement was performed using a high-speed infrared (IR) camera to obtain the surface temperature distribution of the projectile.
The temperature distribution on the spherical surface was reconstructed from the thermographic data, by considering the photoresponse time of the photodetector of the IR camera and the geometric characteristics of the projectile trajectory.
Furthermore, to validate the shock-layer geometry and aerodynamic heating characteristics, a computational fluid dynamics (CFD) simulation was also performed.
The shadowgraph results showed that a detached shock wave and a shock layer were formed in front of the projectile, consistent with the CFD result.
From the thermographic result, it was found that the maximum surface temperature rise during the flight was 24.4 K above the ambient temperature and it decreased with increasing distance from the stagnation point.
The Stanton number distribution was estimated from the reconstructed surface temperature by assuming a one-dimensional transient heat conduction caused during the flight. 
The stagnation Stanton number was calculated to be 0.00366, which was also consistent with both the CFD result and a previously reported empirical correlation.
\end{abstract}
\section{Introduction}
\label{intro}
Accurate prediction of aerodynamic heating is essential for the development of thermal protection systems for hypersonic re-entry vehicles~\citep{uyanna2020thermal} and scramjet inlets~\citep{anderson1989hypersonic,prabhu2021estimation}. 
In the case of the Apollo-shaped capsule re-entry, both field experiments and laboratory experiments \citep{lee1970heat,graves1972apollo,hornung1990role} were conducted to investigate the heat-flux characteristics of the capsule. 
Among laboratory experiments, wind-tunnel testing has been the most conventional technique for investigating the aerodynamic heating characteristics of hypersonic flows.
The Japan Aerospace Exploration Agency (JAXA) developed a free-piston high-enthalpy shock tunnel (HIEST) \citep{itoh2002hypersonic}, capable of producing the stagnation temperatures up to 10,000 K and stagnation pressures up to 150 MPa. 
HIEST has been employed to conduct experiments on the hypersonic aerodyamic-heating characteristics of the Apollo-shaped capsule \citep{tanno2009heat,raespiel2019experimental}. \par

One of the most critical issues in hypersonic re-entry is the laminar-turbulent transition on the capsule surface, which makes the aerodynmic-heating predictions even more difficult.
There have been several studies on the laminar-turbulent boundary transition, necessary to investigate the heat-flux characteristics in a fully-developed turbulent boundary layer.
A prior experimental study using HIEST employed an Apollo-shaped capsule model with boundary-layer tripping, incorporating ``pizza box''-shaped elements to induce forced turbulent transition \citep{tanno2014aeroheating}.
The study reported an increase in the Stanton number St, on the front surface of the capsule in the downstream region of the trip compared to the results obtained from the capsule without tripping.\par

Although wind tunnel tests offer advantages such as lower experimental cost and easier control of test conditions compared with real flight tests, several critical limitations remain.
For example, wind-tunnel tests cannot eliminate the effects of inherent flow disturbances, which are often caused by laminar-turbulent transitions along the tunnel walls.
\cite{parziale2014free} investigated free-stream density perturbations in a shock tunnel at approximately Mach 5.5 using focused laser differential interferometry.
They found that root-mean-square density perturbations were 0.26--6.70\% of the local density depending on cuttoff bandpass wavelengths.
Another critical issue for wind-tunnel tests is the effect on aerodynamic characteristics of a sting used to mount the model.
A previous wind-tunnel test for a free-to-tumble Apollo-shaped capsule model revealed the static and dynamic instabilities of the aerodynamic forces under hypersonic conditions~\citep{moseley1967stability}.
A previous flight test of the hypersonic vehicle X-43A at Mach 5 showed that the pitching moment fluctuated temporally with a frequency of the order of 1 Hz and an amplitude of the order of $10^{-3}$.
Such pitching-moment fluctuations cause angle-of-attack variations in a hypersonic vehicle, which can affect the shock-layer geometry.
This finding implies that wind-tunnel tests, in which the model is fixed by a sting, constrain the flow geometry and are therefore limited in their ability to investigate such dynamic instability effects on aerodynamic heating.\par

Many studies have numerically investigated the aerodynamic heating and flow characteristics of hypersonic re-entry capsules using laminar~\citep{mehta2005numerical,he2025analysis} and Reynolds-Averaged Navier-Stokes (RANS) simulations~\citep{shafeeque2017cfd,rashid2023numerical,inokuma2025numerical}.
However, even numerical studies often face difficulties in ensuring appropriate spatio-temporal resolution and turbulence models capable of capturing laminar-turbulent transition phenomena on the capsule surface.\par

A ballistic range is one option for conducting experiments under free-flight conditions by launching a projectile in quiescent air without a sting.
It can eliminate the freestream-disturbance and model-mounting problems caused in conventional wind tunnels.
Several successful examples have been reported of thermographic measurements of hypersonic projectiles using an Intensified Charge-Coupled Device (ICCD) camera or an infrared (IR) camera in ballistic ranges.
At the NASA Ames Hypervelocity Free Flight Aerodynamic Facility (HFFAF), surface temperatures of hemispherical projectiles at Mach numbers up to 19 were measured, and changes associated with laminar-turbulent transition were investigated~\citep{reda2004aerothermodynamic,reda2010transition,wilder2019rough}.
However, in such methods, the experimental flight speed is limited by the temperature range that the photodetector of the high-speed thermography can detect.
The surface temperature (i.e., radiant intensity) must be sufficiently high for the camera to capture hypersonic motion with a short integration time without motion blur.
In addition, to ensure a high surface temperature, the projectile materials must have low thermal conductivity.
However, such materials often present difficulties in fabrication and surface treatment, which are required to investigate the effects of surface roughness or geometry on laminar-turbulent transition and aerodynamic heating.\par

In the case of \cite{reda2010transition}, the surface temperature of a titanium projectile exceeded 1,500 K at a velocity of 4.5 km/s (Mach 13); this high temperature enabled an integration time of 0.1--1.0 $\mathrm{\mu}$s.
In contrast, Mach numbers (flight velocities) of actual hypersonic flights vary widely; for example, the re-entry velocity ranges from 1.2 to 11 km/s (Mach 3.5--30) at altitudes between 24 and 110 km~\citep{graves1972apollo}, and conceptual hypersonic aircraft operate around Mach 5 \citep{taguchi2012research}.
Therefore, aerodynamic-heating investigations over a wide range of hypersonic velocities are required to develop reliable thermal protection systems.
\cite{wang2019hypersonic} conducted thermographic measurements at Mach numbers between 4.99 and 6.27, and the surface temperature of the aluminum projectile was found to be approximately 340 K, which was much lower than the temperature measured in the HFFAF experiments.
However, such low-temperature measurements limit the available frame rate and resolution, preventing the capture of spatio-temporal temperature variations necessary to obtain accurate heat-flux distributions.
Further investigations of hypersonic flight at various Mach numbers are needed to develop heat-flux compensation methods under experimental conditions with motion blur, in order to overcome such measurement limitations associated with projectile and camera settings.\par

In this study, a heat-flux measurement was conducted for a hypersonic free flight at Mach 5 in a ballistic-range facility, and compensation methods were developed for reconstructing temperature and Stanton number distributions on the projectile surface.
A high-speed IR camera was used to acquire thermographic images.
Unlike conventional wind-tunnel tests, the effects of flow disturbances and sting mounting on temperature measurements could be eliminated.
Stanton number distributions were calculated from the temperature results, compensated by considering the geometrical characteristics of the flight trajectory and the photoresponse time of IR camera photodetector.
The novel compensation methods of this study contributed to extending the range of Mach numbers accessible in ballistic experiments, even under conditions with significant motion blur caused by the long integration time of the IR camera.
The results were compared with computational fluid dynamics (CFD) simulations performed under the same configuration as the experiments.

The primary objective of this study is to establish a generic and reproducible compensation framework for thermographic heat-flux measurements in hypersonic free-flight experiments.
Once the experimental parameters such as projectile geometry, flight velocity, and camera response characteristics are specified, the proposed correction procedure can be systematically applied without reliance on case-specific tuning or operator-dependent judgment.
This feature enables different users and facilities to achieve comparable measurement accuracy under similar experimental conditions, thereby enhancing the general applicability of ballistic-range thermographic measurements.
\section{Methodology}
\subsection{Ballistic range}\label{sec:ballistic range} 
\begin{figure}
\centering
\includegraphics[width=1\linewidth]{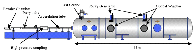}
\caption{Schematic of the ballistic range~(adapted from \cite{numata2008experimental})}
\label{fig:apparatus}
\end{figure}

The experiments were conducted in the ballistic range at the Institute of Fluid Science (IFS), Tohoku University~\citep{numata2008experimental,yamagishi2025three}. 
Figure~\ref{fig:apparatus} shows a schematic of the ballistic range.
In this facility, the projectile was accelerated in an acceleration tube and launched into the 12-m-long test section.
To achieve projectile speeds of up to 2.0 km/s, a single-stage gun operation was employed. A rifle cartridge filled with high-speed smokeless powder was inserted at the end of the launch tube and ignited with a detonator.\par
An aluminum sphere with a diameter $D$ of 8 mm and a mass of 0.72 g was used as the projectile.
It was housed in a four-segment sabot made of polycarbonate (mass: 2.2 g). 
The sphere-sabot assembly was machined into a cylinder 15 mm in diameter and 18 mm long. 
Upon release from a blast remover and exposure to aerodynamic drag forces, the sabot rapidly separated into four pieces.
To ensure complete separation, a sabot stopper, essentially a thick steel plate with a central hole, was placed at the exit of the muzzle blast remover. 
Only the sphere could pass through the hole; the sabot segments were stopped at its front surface.
\subsection{Shadowgraph}\label{sec:shadowgraph} 
\begin{figure}
	\centering
	\includegraphics[width=0.8\linewidth]{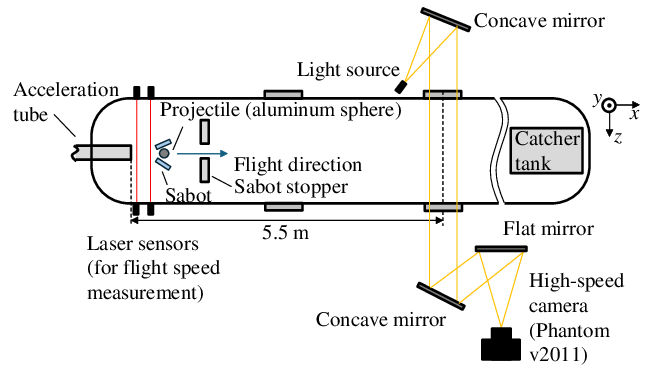}
	\caption{Schematic of the optical setup for the shadowgraph experiment}
	\label{fig:shadowgraph}
\end{figure}

The projectile motion and the shock waves were visualized by shadowgraph imaging, recorded using a high-speed digital video camera (Phantom v2011).
Figure~\ref{fig:shadowgraph} shows a schematic of the optical setup for the shadowgraph experiment. 
A metal halide lamp was used as the light source.
The frame rate and exposure time were 22.0 kHz and 0.43 $\mathrm{\mu}$s, respectively.
The muzzle velocity of the projectile was measured by the time-of-flight method. 
Two continuous 3.2-mW GaAs diode-laser beams (Kikoh Giken Co., Ltd.) were collimated to approximately 1 mm in diameter and positioned 200 mm apart in parallel near the muzzle exit.
Sudden interruptions of the laser beams by the passing projectile were detected by PIN diodes, and the muzzle velocity was estimated from the measured time interval.
The output signals were recorded on a digital memory (Yokogawa DL750), with timing errors less than 1 $\mathrm{\mu}$s.\par
The center of the observation window was located 5.5 m downstream from the muzzle exit (i.e., the exit of the acceleration tube).
The projectile velocity slightly decreased owing to aerodynamic drag, with an attenuation rate of approximately 11\% between the muzzle exit and the entrance to the cryogenic chamber~\citep{numata2008experimental}.
The output signal from the time-of-flight measurement triggered both the high-speed video camera and the light source.
\subsection{Thermography}\label{sec:thermography} 
\begin{figure}
	\centering
	\includegraphics[width=0.8\linewidth]{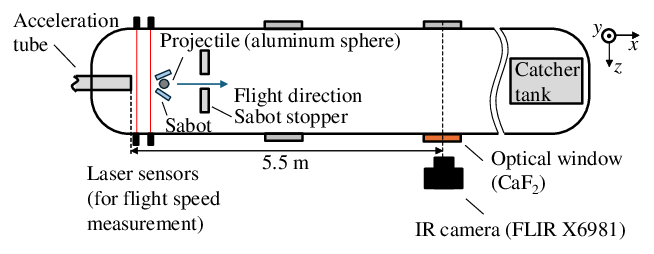}
	\caption{Schematic of the experimental apparatus for the thermography measurement}
	\label{fig:IR}
\end{figure}

A thermographic technique was applied to measure the surface-temperature distribution of the projectile.
Figure~\ref{fig:IR} shows a schematic of the experimental apparatus for thermography.
To capture hypersonic aerodynamic heating, a mid-wave high-speed IR camera (FLIR X6981) was used.
Emissivity compensation was performed for the measured temperature data following the method of \cite{Tran2017effects}.
The calibrated temperature range was set to 283--363 K to match the expected projectile-surface temperature range.
Within this range, the frame rate was set to its maximum value (2.03 kHz), corresponding to an integration time of 0.479 ms.
The noise equivalent temperature difference (NETD) was 0.02 K.
The observation window was located at the same position as in the shadowgraph experiment, 5.5 m from the muzzle exit.
For the spectral range of the thermography by the IR camera (3--5 $\mathrm{\mu}$m), a $\mathrm{CaF_2}$ window (Edmund 23-525, diameter: 75 mm, thickness 5 mm) was used as the optical window and the IR camera was placed directly in front of the window.

\subsection{Experimental conditions} \label{sec:experimental conditions} 
Table~\ref{tab:experimental conditions} shows the experimental conditions for the shadowgraph and thermography measurements. 
The experiments were conducted under ambient conditions (ambient pressure of $p_\infty$ and ambient temperature of $T_\infty$). 
The measured muzzle velocities $U_\infty$ were 1.64--1.69 km/s, corresponding to the Mach number $\mathrm{M_a}$ (= $U_\infty/a_\infty$) of 4.76--4.90, where $a_\infty$ is the speed of sound at the temperature $T_\infty$. 
The stagnation enthalpies $H_0$ ($= C_p T_\infty+ (1/2)U_\infty^2)$ were 1.64--1.73 MJ/kg, where $C_p$ is the specific heat at constant pressure.

\begin{table}
	\caption{Experimental conditions for the shadowgraph and thermography measurements}
	\label{tab:experimental conditions}       
	\begin{tabular}{c|c|c}
		\hline
		&Shadowgraph &Thermography \\
		\hline\hline   
		Ambient pressure, $p_\infty$ [kPa] &101 & 101 \\
		\hline
		Ambient temperature, $T_\infty$ [K] &298 & 295 \\
		\hline
	    Muzzle velocity, $U_\infty$ [km/s] &1.69 & 1.64 \\
		\hline
		Mach number, $\mathrm{M_a}$ [-] &4.90 & 4.76 \\
		\hline
	    Stagnation enthalpy, $H_0$ [MJ/kg] &1.73 & 1.64 \\
		\hline
	\end{tabular}
\end{table}

Table~\ref{tab:camera_settings} summarizes the camera settings for both the shadowgraph and thermography experiments.
The exposure time for the shadowgraph was 0.430 $\mathrm{\mu}$s, during which the projectile moved approximately 0.727 mm (9\% of the projectile diameter, $D$ = 8 mm). 
In contrast, the integration time for the thermography (equivalent to the exposure time) was 0.479 ms, during which the projectile was expected to move along a distance of 785 mm (approximately 98 times its diameter), causing motion blur.
A compensation technique was therefore developed to reconstruct thermographic data affected by motion blur, as discussed below.
\begin{table}
	\caption{Camera settings for the shadowgraph and thermography experiments}
	\label{tab:camera_settings}       
	\begin{tabular}{c|c|c}
		\hline
		&Shadowgraph &Thermography \\
		\hline\hline
		Camera & High-speed camera&  Midwave high-speed IR camera\\
		& (Phantom v2011)& (FLIR x6981)\\
		\hline
		Frame rate &22.0 kHz & 2.03 kHz\\
		\hline
		Exposure/integration time &0.430 $\mathrm{\mu}$s  &0.479 ms\\
		\hline
		No. pixels &1280$\times$800& 612$\times$128\\
		\hline
		
	\end{tabular}
\end{table}
\subsection{Numerical method}\label{sec:cfd} 
\begin{figure}
	\centering
	\includegraphics[width=0.8\linewidth]{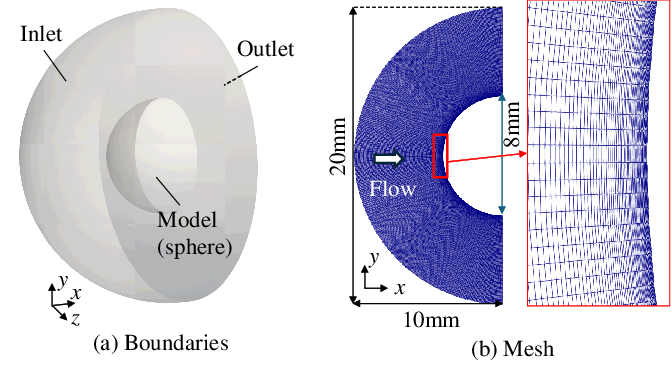}
	\caption{Schematics of the computational domain: (a) Boundary geometries and (b) central cross-sectional view of the computational mesh}
	\label{fig:computational_domain}
\end{figure}

A laminar CFD simulation was also performed to validate the experimental results. 
Figure ~\ref{fig:computational_domain} shows the computational domain schematics.
A spherical model, with a diameter $D$ of 8 mm, was utilized, identical to that in the experiments. 
A Cartesian coordinate system was defined with $x$, $y$, and $z$ axes, with the origin located at the front edge of the spherical surface.
The computational domain consisted of an ``Inlet'' surface with a Dirichlet boundary condition, an ``Outlet'' surface with a Neumann boundary condition, and a ``Model (sphere)'' surface with a no-slip velocity condition, as shown in Fig.~\ref{fig:computational_domain}(a). 
The freestream Mach number $\mathrm{M_{\infty}}$ was set to 4.76; the Reynolds number based on $D$, $\mathrm{Re}$ = $U_\infty D/\nu$, was $\mathrm{Re} = 8.45\times10^5$, where $U_\infty$  is the freestream velocity and $\nu$ is the kinematic viscosity. 
The wall temperature $T _\mathrm{wall}$ for the capsule front surface was 300 K, and the stagnation enthalpy $H _0$ was 1.64 MJ/kg. These parameters were consistent with the experimental conditions. \par

The compressible continuity, Navier--Stokes, and energy equations were solved using the finite volume method implemented in OpenFOAM-v2312 (solver: rhoCentralFoam) with a second-order central spatial scheme with the vanLeer limiter and a first-order Euler temporal scheme. Structured mesh was used as shown in Fig.~\ref{fig:computational_domain}(b). The wall-surface averaged value for the first near-wall resolution in wall units, $\Delta^+$, was 6.22 and the total cell count was 9,751,995.

\section{Results and discussions}
\subsection{Shadowgraph results}
\begin{figure}
	\centering
	\includegraphics[width=0.8\linewidth]{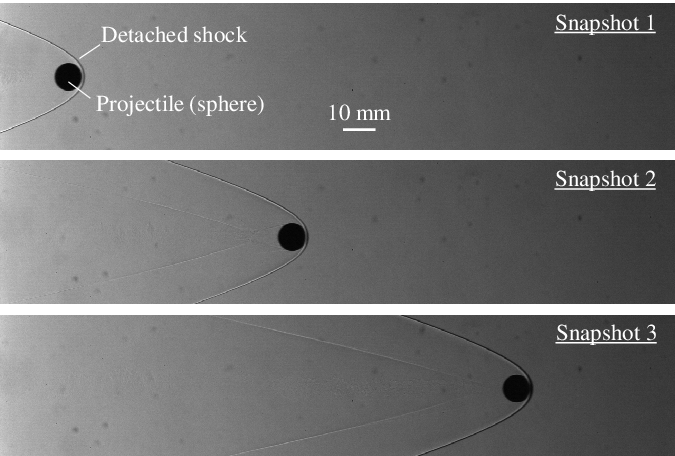}
	\caption{Shadowgraph snapshots of the projectile in flight; the time interval between frames was 45 $\mathrm{\mu}$s}
	\label{fig:shadowgraph_result}
\end{figure}

Figure~\ref{fig:shadowgraph_result} shows shadowgraph snapshots of the projectile in flight, moving from left to right.
The formation of a detached shock wave ahead of the projectile was clearly observed in each snapshot. 
The time interval between frames was 45 $\mathrm{\mu}$s, and the estimated flight speed was 1.39 km/s, which decreased from the muzzle velocity by aerodynamic drag.
\par
\begin{figure}
	\centering
	\includegraphics[width=0.6\linewidth]{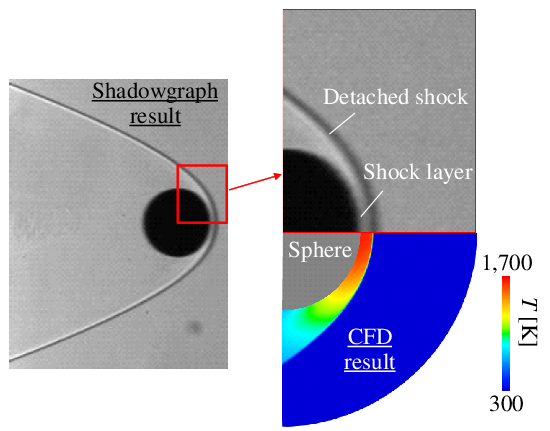}
	\caption{Comparison between the shadowgraph (top right, enlarged view of the left image) and CFD (bottom right) results of the detached shock wave; in the CFD result, colors represent temperature}
	\label{fig:comparison}
\end{figure}

Figure~\ref{fig:comparison} compares the experimental and CFD results of the detached shock wave in front of the projectile.
The CFD temperature distribution around the sphere revealed a high-temperature region in the shock layer caused by rapid compression behind the detached shock wave, which induced aerodynamic heating.
\cite{yamagishi2025three} conducted a background-oriented schlieren (BOS) measurement in the ballistic range facility to obtain the density distribution of the shock layer ahead of the projectile.
Although their flight Mach number (Mach 1.20) was much lower than that of the present study, a high-density region (indicating strong compression) was observed, consistent with the current CFD results.
A comparison between the present shadowgraph and CFD results demonstrated good agreement in the geometry of the detached shock wave, indicating that the high-temperature region due to shock compression was also formed in the present experiments.
\subsection{Thermographic results}
\subsubsection{Temperature dirstibution}
\begin{figure}
	\centering
	\includegraphics[width=1\linewidth]{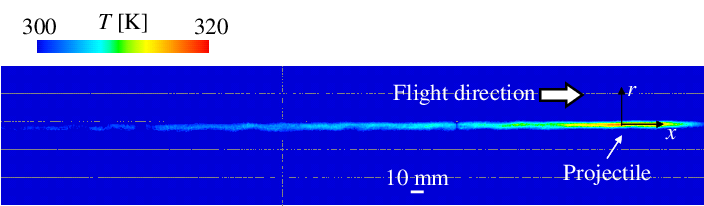}
	\caption{Thermographic result showing the raw temperature distribution, $T=T_\mathrm{raw}$} 
	\label{fig:2D_temperature}
\end{figure}

\begin{figure}
	\centering
	\includegraphics[width=0.6\linewidth]{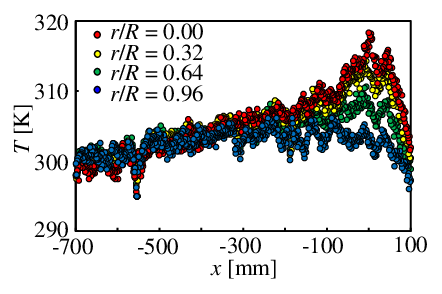}
	\caption{Temperature distributions of $T=T_\mathrm{raw}$ plotted along the flight direction $x$ for each normalized radial position $r/R$ on the projectile surface} 
	\label{fig:1D_temperature}
\end{figure}

Figure~\ref{fig:2D_temperature} shows the raw thermographic data of temperature, $T=T_\mathrm{raw}$.
During the integration time of 0.479 ms, the projectile traveled approximately 785 mm, resulting in a long ellipsoidal structure in the image caused by motion blur from the photodetector of the IR camera. 
The thermographic image shows that temperature decreased from the peak (red region) with increasing radial distance $r$ and axial distance $x$.
From the front of the projectile to the peak position, the temperature rose during the time response process after the photodetector detected the projectile temperature. On the other hand, from the peak position to the rear, the temperature gradually changed from the projectile temperature to the ambient temperature during the time from when the projectile passed that position and was detected by the photodetector until the shutter closed. As a result, the temperature decreased in the negative $x$ direction.
Because the projectile was flying at approximately Mach 5, the radial displacement during the observation period was negligible, as shown in Fig.~\ref{fig:shadowgraph_result}.
Therefore, the temperature distribution in the $r$ direction was assumed to be unaffected by motion blur.
In contrast, the $x$ directional (flight-directional) temperature distribution was influenced by the finite rise time of the photodetector during the integration time of the camera.
Figure~\ref{fig:1D_temperature} shows the raw data, $T=T_\mathrm{raw}$, extracted from Fig.~\ref{fig:2D_temperature} along the projectile path and plotted against the flight-directional coordinate, $x$.
The origin of $x$ is the peak position of $T_\mathrm{raw}$.
With an IR camera resolution of 1.28 mm/pixel, the spherical projectile of radius $R$ of 4 mm was resolved with four pixels ($r/R$ = 0.00, 0.32, 0.64, and 0.96) in the radial direction.
The temperature increased from the right edge of the frame to the peak position ($x$ = 100--0 mm), corresponding to the photoresponse of the photodetector from ambient temperature to the projectile surface temperature.
In this study, motion-blur compensation was carried out by focusing on this temperature-rise process from the front of the model to the peak temperature observed in the thermographic image.
\par
\begin{figure}
	\centering
	\includegraphics[width=0.6\linewidth]{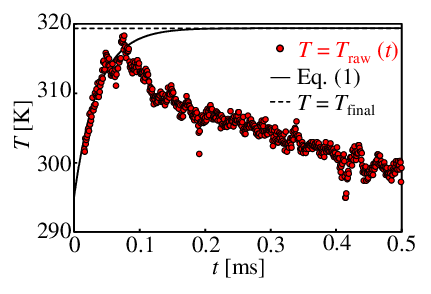}
	\caption{Fitting of the temperature response given by Eq. (\ref{eq:lu}) (black dashed line) to the measured temperature rise $T=T_\mathrm{raw}$ at $r/R$ = 0.00 (red solid circles)} 
	\label{fig:fitting_T_raw}
\end{figure}

The dynamic response of the photodetector current $I$ follows the equation, $I=I_0\left[1-\mathrm{exp}(-t/\tau)\right]$, where $I_0$, $t$, and $\tau$ are the steady-state current, time, and rise time, respectively~\citep{lu2015gas-dependent}.
\cite{li2017experimental} used a cubic function to calibrate the relationship between temperature and digital counts (current counts) of an IR camera in a 290--370 K range. 
Because the temperature range in the present study was 293--323 K, a linear relationship between temperature and current was assumed, yielding
\begin{equation}
	T-T_\infty=(T_\mathrm{final}-T_\infty)\left[1-\mathrm{exp}\left(-\frac{t}{\tau}\right)\right],
	\label{eq:lu}
\end{equation}
where $T_\mathrm{final}$ and $T_\infty$ denote the final temperature of the photo response and the initial temperature (ambient temperature), respectively.
Figure~\ref{fig:fitting_T_raw} shows the fitting of Eq. (\ref{eq:lu}) to the measured $T=T_\mathrm{raw}(t)$ at $r/R=0.00$.
The peak value of $T_\mathrm{raw}$ was observed at approximately $t$ = 0.08 ms. 
The fitting procedure was applied to the temperature-rise period after the projectile surface temperature was detected ($t$ $<$ 0.05 ms).
The uncertainty in the determination of $T_\mathrm{final}$ was within 0.4 K when the fitting time range was varied between 0.05 and 0.08 ms.
From the fitting results, the rise time $\tau$ was determined to be 36 $\mu$s. 
The spatial distribution of $T_\mathrm{raw}(x)$ was then converted into a temporal profile $T_\mathrm{raw}(t)$ using the relation $x=U_\infty(t-\mathrm{\Delta} t)$ , where $t$ is the time between the moment the projectile passes position $x$ and the moment the IR camera shutter closes.
Here, $t=0$ corresponds to the start of the photodetector response, and $\mathrm{\Delta} t$ represents the time difference between when the projectile exits the frame and when the shutter closes.\par
\begin{figure}
	\centering
	\includegraphics[width=0.6\linewidth]{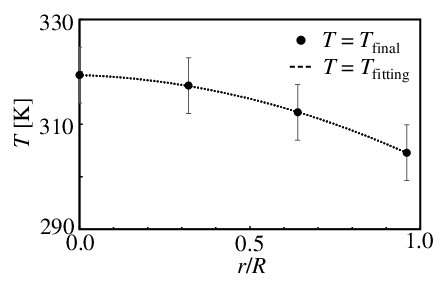}
	\caption{Temperature distribution, $T = T_\mathrm{final}$, plotted against $r/R$ (black solid circles) with its quadratic fitting curve, $T=T_\mathrm{fitting}=-13.9(r/R)^2-2.11r/R+319$ [K] (black broken line)} 
	\label{fig:fitting}
\end{figure}

\begin{figure}
	\centering
	\includegraphics[width=0.6\linewidth]{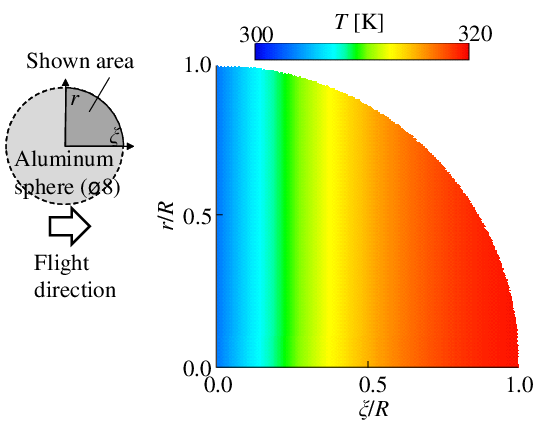}
	\caption{Reconstructed temperature distribution on the projectile surface, based on $T=T_\mathrm{fitting}$} 
	\label{fig:Reconstructed_temperature}
\end{figure}

Figure~\ref{fig:fitting} shows the corrected temperature distribution, $T = T_\mathrm{final}$, obtained from Eq. (\ref{eq:lu}). 
The maximum temperature rise above ambient temperature due to aerodynamic heating was 24.4 K at the stagnation point.
The error bars show the measurement uncertainties at 95\% coverage, which is addressed in Section \ref{sec:uncertainty}.
The temperature distribution was interpolated using a quadratic fitting curve, $T=T_\mathrm{fitting}=-13.9(r/R)^2-2.11r/R+319$ [K].
Because the projectile trajectory was parallel to the $x$-axis, the temperature distribution was reconstructed by projecting $T_\mathrm{fitting}(r)$ onto the spherical surface. 
Figure~\ref{fig:Reconstructed_temperature} shows the reconstructed surface temperature distribution based on $T=T_\mathrm{fitting}$. 
Here, $\xi$ denotes the flight-directional distance from the projectile center.
The maximum temperature occurred at the stagnation point ($\xi/R$, $r/R$) = (1.0, 0.0), decreased monotonically toward $\xi/R$ = 0.0.
\subsubsection{Stanton number distribution}
\begin{figure}
	\centering
	\includegraphics[width=0.6\linewidth]{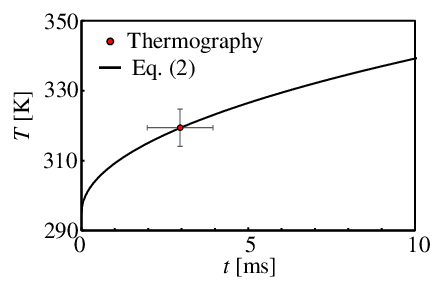}
	\caption{Transient temperature predicted by Eq. (\ref{eq:heat_transfer}) fitted for the thermographic result of $T=T_\mathrm{final}$ at $r/R=0.00$} 
	\label{fig:st_calc}
\end{figure}

\begin{figure}
	\centering
	\includegraphics[width=0.6\linewidth]{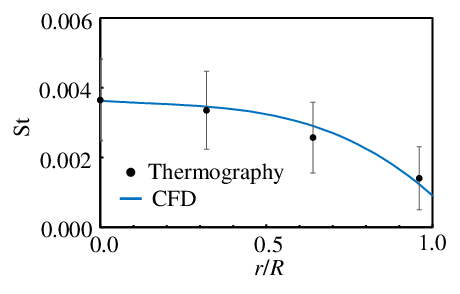}
	\caption{Stanton number $\mathrm{St}$ obtained from thermography (black solid circles) and CFD (blue solid line) plotted against $r/R$}
	\label{fig:stanton_number_comparison}
\end{figure}

The Stanton number distribution was estimated from the temperature profile, $T_\mathrm{final}$.
Because of uncertainty in the timing of sabot separation, the onset of aerodynamic heating may have ranged from the muzzle exit to the sabot stopper. 
Moreover, the projectile velocity decreased from 1.64 km/s (at the muzzle exit) to 1.39 km/s (at the optical window) due to aerodynamic drag.
Thus, the heating duration during which the projectile was exposed to aerodynamic heating was estimated to range from 1.97 to 3.95 ms, which yields an uncertainty of 33\% for the heating duration. The heating time was determined to be 2.96 ms as an average of 1.95 and 3.95 ms.
Here, the temperature penetration depth ($\Delta = (12at)^{0.5}$, $a$: thermal diffusivity of the projectile, $t$: heating time)~\citep{jsme1989heat} estimated for this heating time was approximately 1.9 mm, which was sufficiently thinner than the projectile diameter of 8 mm.
Therefore, the analytical solution of one-dimensional transient heat conduction in a semi-infinite solid was used for estimating the temporal temperature variation due to aerodynamic heating as follows~\citep{carslan1956conduction}:
\begin{equation}
	\frac{T-T_{\infty}}{T_0-T_{\infty}}=1-\mathrm{exp}\left(\frac{h^2at}{\lambda^2}\right)\mathrm{erfc}\left(\frac{h\sqrt{at}}{\lambda}\right),
	\label{eq:heat_transfer}
\end{equation}
where $T_0$, $h$, and $\lambda$ denote the stagnation temperature, heat-transfer coefficient, and thermal conductivity of the projectile, respectively.
The heat-transfer coefficient $h$ in Eq. (\ref{eq:heat_transfer}) was determined by fitting the thermographic result of $T_\mathrm{final}$ as shown in Fig.~\ref{fig:st_calc}. The uncertainty for the heating-time estimation is indicated by the horizontal error bars.
From the fitted $h$ value, the Stanton number $\mathrm{St}$ was calculated as
\begin{equation}
	\mathrm{St}=\frac{h}{\rho_\infty C_p U_\infty},
	\label{eq:stanton_number}
\end{equation}
where $\rho_\infty$ is the ambient density.
Figure~\ref{fig:stanton_number_comparison} shows the obtained $\mathrm{St}$ values using Eqs. (\ref{eq:heat_transfer}) and (\ref{eq:stanton_number}). 
The error bars show the measurement uncertainties at 95\% coverage, which is addressed in Section \ref{sec:uncertainty}.
The thermographic St results agreed closely with the CFD result, validating the proposed thermographic heat-flux measurement method.
The stagnation Stanton number $\mathrm{St_s}$ was also compared with the empirical correlation for a sphere proposed by~\cite{Zhou2023aerothermoal}: 
\begin{equation}
	\mathrm{St_s}=2.60\mathrm{Ma}^{0.5}\mathrm{Re}_R^{-0.5}F, F=\frac{H_0-H_\infty}{H_0}\frac{1}{T_\infty^{0.085}},
	\label{eq:zhou}
\end{equation}
where $\mathrm{Re}_R=U_\infty R/\nu$ is the Reynolds number based on $R$ ($\nu$: kinematic viscosity), and $H_\infty$ is the freestream enthalpy.
Table~\ref{tab:St} lists the $\mathrm{St_s}$ values for the thermography, CFD, and empirical correlation.
The result obtained from the empirical correlation was 25.1\% higher than those from the thermography and CFD results.
This discrepancy may be attributed to the experimental conditions under which the empirical correlation was derived, because the heat-flux characteristics can differ depending on whether the flow is turbulent or laminar (the present thermography and CFD analyses were conducted under laminar conditions).
Nevertheless, the difference was within the experimental measurement uncertainty (31.8\%), and all three results showed close overall agreement.\par
\begin{figure}
	\centering
	\includegraphics[width=0.6\linewidth]{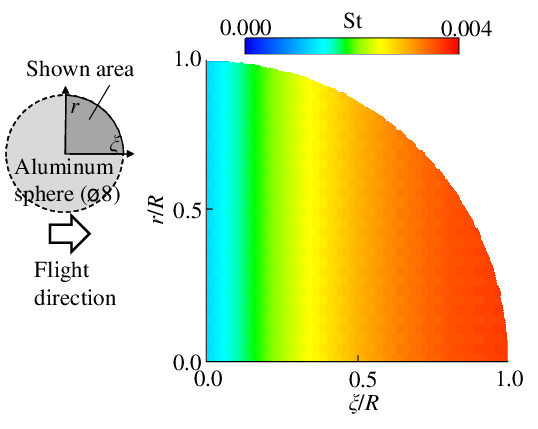}
	\caption{Reconstructed distribution of the Stanton number $\mathrm{St}$ on the projectile surface}
	\label{fig:stanton_number}
\end{figure}

\begin{table}
	\caption{Comparison of the stagnation Stanton numbers $\mathrm{St_s}$ obtained from thermography, CFD, and the empirical correlation by \cite{Zhou2023aerothermoal}}
	\label{tab:St}       
	\begin{tabular}{c|c|c|c}
		\hline
		&Thermography &CFD &Zhou et al. 2023\\
		\hline\hline   
		$\mathrm{St_s}$ &0.00366 &0.00363 &0.00458\\
		\hline
	\end{tabular}
\end{table}
Similar to the temperature distribution, the Stanton number was reconstructed on the spherical surface by applying a quadratic fitting, $\mathrm{St}=-2.12\times10^{-3}(r/R)^2-3.09\times10^{-4}r/R+0.00366$, for the thermographic result shown in Fig.~\ref{fig:stanton_number_comparison}.
Figure \ref{fig:stanton_number} shows the Stanton number profile over the spherical surface. 
The maximum value occurred at the stagnation point ($\xi/R$, $r/R$) = (1.0, 0.0) and decreased monotonically toward $\xi/R$ = 0.0. 
Therefore, the reconstructed St distribution was successfully obtained using the proposed compensation methods, which account for the geometry of the projectile trajectory, the temporal response of the photodetector, and data interpolation. 
These techniques demonstrate strong potential for overcoming the experimental limitations of hypersonic aerodynamic heating research, particularly regarding spatio-temporal resolution and motion-blur constraints.
Although the present study was based on a single thermographic dataset, the validity of the proposed method was supported by multiple layers of physical consistency.
The reconstructed temperature and Stanton number distributions were derived from established physical models, including the photodetector response function and one-dimensional transient heat conduction theory.
Furthermore, the uncertainty propagation analysis quantitatively demonstrated that the obtained heat-transfer characteristics remained robust within the estimated measurement uncertainties.
The close agreement with both CFD results and an independent empirical correlation further supported the physical soundness of the proposed approach, even with a limited number of experimental samples.
\subsubsection{Measurement uncertainty}\label{sec:uncertainty} 
The measurement uncertainties with a 95\% confidence level were evaluated for $T_\mathrm{final}$ and $\mathrm{St}$ following the method described in~\cite{ansi1985measurement}.
The uncertainty of $T_\mathrm{final}$ at the stagnation point was estimated to be $\pm$5.32~K (21.8\% of the temperature rise, $T_\mathrm{final} - T_\infty$), which mainly originated from the uncertainties in the IR camera temperature measurement, emissivity correction, and compensation using Eq.~(\ref{eq:lu}).
The uncertainty of $\mathrm{St_s}$ was $\pm$0.00116 (31.8\%), primarily due to the uncertainties in $\rho_\infty$, $U_\infty$, and the determination of $h$ using Eq.~(\ref{eq:heat_transfer}).

\section{Conclusions}
\label{sec:conclusions}
In this study, novel thermographic measurement and compensation techniques were developed and demonstrated for acquiring the heat flux of a hypersonic projectile in ballistic flight.
The surface temperature distribution of a spherical projectile was successfully reconstructed from high-speed infrared (IR) thermography by accounting for the geometry of the flight trajectory and the photo-response characteristics of the IR camera.

The shadowgraph results revealed the presence of a detached shock wave and a well-defined shock layer in front of the projectile, consistent with the computational fluid dynamics (CFD) simulation.
The reconstructed temperature and Stanton number distributions agreed closely with both the CFD result and empirical correlation, confirming the validity and accuracy of the proposed reconstruction method.

The developed technique enables non-intrusive measurement of transient heat transfer in free-flight hypersonic experiments, even at Mach numbers where motion blur typically occurs.
This approach is expected to contribute significantly to future research on aerodynamic heating and thermal protection systems for hypersonic vehicles.
\section*{Acknowledgements}
We thank Mr. T. Goto, Mr. Y. Onodera (Chino Corporation), and Mr. T. Takaya (Naruse Corporation) for their kind instruction and advice for thermography. We thank Prof. H. Nagai, Mr. T. Ogawa (Tohoku University), Prof. M. Furudate, Mr. Y. J. Kim (Chungnam National University), Prof. B. J. Lee, and Mr. J. Kim (Seoul National University) for their help and valuable comments to this work. We thank Mr. B. Ridolfi (Polytech Lyon) for his support in the CFD of this work. The CFD of this work was performed using the AFI-NITY II super computer at the Institute of Fluid Science, Tohoku University.
%

\section*{Funding} 
Part of this work was supported by JST CREST (Grant No. JPMJCR24Q6).

\section*{Conflict of interest}
The authors declare that they have no conflict of interest.

\section*{Data availability} 
The data supporting the findings reported in this paper are shadowgraphy and thermographic images. They are available from the authors upon request.

\section*{Author contributions} 
K.I. contributed to the conceptualization, investigation, methodology, data curation, analysis, validation, and original draft preparation. A.Y. contributed to the conceptualization, funding acquisition, review, and editing. Y.W. contributed to the methodology, data curation, review, and editing. K.O. contributed to the technical expertise, review and editing.

\bibliographystyle{spbasic}      

\bibliography{inokuma_bib}   


\end{document}